\documentclass{ECOC-author}
\usepackage{amsmath,amssymb,amscd,latexsym,dsfont}
\usepackage{float}
\usepackage{comment}
\usepackage{color}
\usepackage{graphicx}
\usepackage{subfigure}
\usepackage{enumerate}
\usepackage{stfloats}
\usepackage{psfrag}
\usepackage{setspace}
\usepackage{balance}
\usepackage{tikz}
\usepackage{pgfplots,amsfonts}
\usepackage{pgfplotstable}
\usepackage{bm}
\usetikzlibrary{shapes}
\usetikzlibrary{spy}
\usetikzlibrary{fit}
\usetikzlibrary{shapes.multipart}
\usetikzlibrary{positioning}

 \usetikzlibrary{external}
 \tikzsetfigurename{tikz_compiled} 

\pgfdeclarelayer{background}
\pgfsetlayers{background,main} 
\usepackage{algorithmic}
\definecolor{myDarkGreen}{rgb}{0.00000,0.58824,0.00000}%

\pgfplotsset{compat=newest}

\usepackage{stfloats}

\usetikzlibrary{plotmarks,matrix,chains,scopes,fit,calc,shapes,positioning,decorations,intersections,arrows,backgrounds,shadows}

\newlength\FigureHeight
\newlength\FigureWidth
\definecolor{myDarkGreen}{rgb}{0.00000,0.58824,0.00000}%
\definecolor{uniform}{rgb}{0.00000,0.58824,0.00000}%
\definecolor{AWGNreference}{rgb}{0.00000,0.58824,0.00000}%

\newcommand{\bX}{\boldsymbol{X}}

\newcommand{\bY}{\boldsymbol{Y}}\newcommand{\by}{\boldsymbol{y}}

\newcommand{\bs}{\boldsymbol{s}}

\begin{document}

\title{FOUR-DIMENSIONAL POLARISATION-RING-SWITCHING FOR DISPERSION-MANAGED OPTICAL FIBRE SYSTEMS}

 \author{Bin Chen\textsuperscript{1,2}, Chigo Okonkwo\textsuperscript{1},  Hartmut Hafermann\textsuperscript{3}, Alex Alvarado\textsuperscript{1}}

\address{\textsuperscript{1}Department of Electrical Engineering, Eindhoven University of Technology, The Netherlands.  {{\underline{b.c.chen@tue.nl}}}\\
\textsuperscript{2}School of Computer and Information, Hefei University of Technology, China.
\\
  \textsuperscript{3}Mathematical and Algorithmic Sciences Lab, Paris Research Center, Huawei Technologies France SASU, France.\\
}


\begin{abstract}
The recently introduced 4D 64-ary polarisation-ring-switching format is investigated in dispersion-managed systems. Numerical simulations show a reach increase of $25\%$ with respect to PM-8QAM. This gain is achieved from the nonlinear tolerance of the format and a 4D demapper using correlated noise assumptions.
\end{abstract}
\maketitle
\section{Introduction}
\vspace{-0.5em}
Fibre nonlinearities are considered to be one of the limiting factors for achieving higher information rates in coherent optical fibre transmission systems \cite{EssiambreJLT2010,SecondiniJLT2013}.  Multiple digital signal processing (DSP) algorithms have been investigated to reduce the impact of intra-channel and inter-channel nonlinearities \cite{Cartledge:17}.  One appealing alternative is to properly design multi-dimensional (MD) modulation formats, which can minimize the signal power variations, and thus, reduce the nonlinear interference (NLIN) power. This property is especially suitable for dispersion-managed (DM) links with a strong nonlinear cross-phase modulation (XPM) \cite{Chagnon:13,ReimerOFC2016,Kojima2017JLT}. 

We recently proposed in \cite{BinChenJLT2019} the 4D 64-ary polarisation-ring-switching (4D-64PRS) format, which was obtained by jointly optimising the coordinates and labeling of the constellation. 4D-64PRS is a nonlinearity-tolerant format which was shown to outperform other formats at the same spectral efficiency (6 bit/4D-sym) in both the linear and nonlinear channels. All the  reported results in \cite{BinChenJLT2019} were obtained for dispersion unmanaged systems only. The first contribution of this paper is to extend those results to DM links.

In this paper, we use generalised mutual information (GMI) as a figure of merit. GMI is very popular because it directly shows the net data rates obtained by state-of-the-art soft-decision forward error correction (SD-FEC) \cite{AlvaradoJLT2018}. 
GMI calculations can include the demapper structure at the receiver. In this case, the GMI is calculated using log-likelihood ratios (LLRs, or soft bits), which are in turn calculated using a particular assumption on the channel. 

Typically, circularly symmetric Gaussian statistics are assumed in the LLR calculation. This assumption is accurate for uncompensated long-haul fibre systems \cite{ErikssonJLT2016,KeykhosraviARXIV2018}.
However, in systems with strong phase noise (e.g., due to residual  phase noise) or large nonlinear fibre effects (like in a DM link), this assumption is no longer valid. Transmission experiments  have shown a decrease in achievable information rates and coding gain if there is a strong mismatch between the true channel and the channel assumption made by the demapper \cite{ZhaoOFC2013,ErikssonJLT2016}. The second contribution of this paper is to show that relatively large GMI gains are obtained when 4D-64PRS is combined with a demapper that lifts the circularly symmetric Gaussian assumption.

In this paper, we show the nonlinearity tolerance of 4D-64PRS over PM-8QAM in DM links and a demapper that treats the four dimensions jointly and allow correlations between them. It is shown that 4D-64PRS outperforms PM-8QAM by up to  $0.55$ bit/4D-sym in terms of GMI ($12\%$ rate increase) for a 2400~km DM transmission link. 4D-64PRS also yields 600 km ($25\%$) reach increase. Finally, we also show the gain increases as the number of WDM channels increase. This leads us to believe that 4D-64PRS is well-suited for ultra wideband fibre systems.

\vspace{-0.3em}
\section{The 4D-64PRS Modulation Format}
\begin{figure}[!b]
	\vspace{-1.5em}
	\centering
	\includegraphics[width=0.23\textwidth]{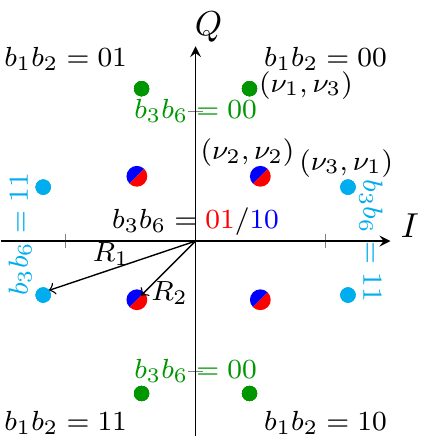}
	\includegraphics[width=0.23\textwidth]{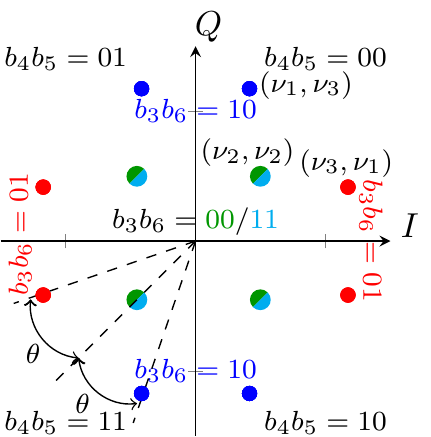}
	\caption{2D-projections of the designed 4D-64PRS modulation and associated bits/polarisation bits mapping. The rings are given by $R_1^2={\nu_1^2+\nu_3^2}$ and $R_2^2=2\nu_2^2$.}
	\label{fig:4D_64_modulation_label}
\end{figure}
Fig.~\ref{fig:4D_64_modulation_label} shows the coordinates and the corresponding binary labeling of  the 4D-64PRS with $M=2^m=64$ points (6~bit/4D-sym) for two 2D-projections. 
The coordinates of 4D-64PRS are such
\begin{align}\label{eq:alphabet}\nonumber
\vspace{-1.5em}
\bs_i \in &\{[\pm \nu_1, \pm \nu_3, \pm \nu_2, \pm \nu_2], [\pm \nu_3, \pm \nu_1, \pm \nu_2, \pm \nu_2],\\ \nonumber&
[\pm \nu_2, \pm \nu_2, \pm \nu_1, \pm \nu_3], [\pm \nu_2, \pm \nu_2, \pm \nu_3, \pm \nu_1]\},
\vspace{-1em}
\end{align}
where $i\in\{1,2,\cdots,64\}$,  $\nu_2=\operatorname{Re}\left(R_1e^{j(\phi_1)}\right) =\sqrt{1/2}R_2$, $\nu_1= \operatorname{Re}\left(R_1e^{j(\phi_2)}\right) $ and $\nu_3= \operatorname{Im}\left(R_1e^{j(\phi_2)}\right)$, $R_j$ and  $\phi_j$ indicate the radius and phase of inner ($j=1$) and outer ring ($j=2$).
$\theta=\phi_1-\phi_2>0$ is the relative phase between inner  and outer ring points.
We can observe that the 4D-64PRS is highly symmetric.
The quadrants in each polarisation (or orthants in 4D space) are defined by 4 out of 6 bits, namely, $[b_1,b_2]$ and $[b_4,b_5]$ for the first and second polarisation, respectively. These $4$ bits can be seen as the ones determining which of the 16 orthants is to be chosen. The remaining 2 bits [$b_3,b_6$] determine the symbol from the 4 possible constellation points in the same orthant.
In other words, [$b_3,b_6$] determine the color of the transmitted points  while [$b_1,b_2,b_4,b_5$] determine the coordinate of the point in the same color through $[(-1)^{b_2}s_{i,1},(-1)^{b_1}s_{i,2},(-1)^{b_4}s_{i,3},(-1)^{b_3}s_{i,4}]$ with $i\in\{1,2,\ldots,64\}$.

  \vspace{-0.2em}
\section{Nonlinear Channel Performance}
 \vspace{-0.5em}
The simulation setup is as follows: 
a dual-polarisation multi-span WDM system with 11 co-propagating channels are generated at the transmitter at a symbol rate of 45 GBaud with a root-raised cosine (RRC) pulse shape with a roll-off factor of 0.1.
Each WDM channel carries  $2^{16}$ symbols in two polarisations at the same power. The WDM optical signals are launched into a multi-span optical link. Each span consists of a 80~km  large effective area fibre (LEAF) with {0.219 dB/km attenuation, 4.255 ps/nm/km dispersion and 1.464 (w$\cdot$km)$^{-1}$ nonlinear coefficient}. Each span is followed by an ideal $100\%$ in-line chromatic dispersion compensation fibre and an erbium-doped fibre amplifier (EDFA) with a noise figure of 5 dB.
The transmission link is simulated employing a split-step Fourier solution of the Manakov equation with a step size of 0.1~km to guarantee the accuracy.
polarisation mode dispersion (PMD) is not considered. 

After match filter and ideal phase compensation, performance of the constellation with cardinality $M = 2^m$ is evaluated via the GMI 
as   \cite{AlvaradoJLT2018}
\begin{align}
\vspace{-1em}
\text{GMI}
\approx  m-\frac{1}{N_s}\sum_{k=1}^{m}\sum_{j=1}^{N_s}\log_2\left(1+e^{(-1)^{b_{k,j}L_{k,j}}}\right),
\vspace{-1em}
\end{align}
where $L_{k,j}$ is the log-likelihood ratios  (LLRs) with $k$ denoting the bit position and $j$ denoting
the $j$th received symbol with $j=1,2,\cdots,N_s$.
For the general case of an $N$-dimensional demapper, the LLRs are calculated by the demapper using the channel law
$f_{\bY|\bX}$. 
\begin{equation}\label{eq:llr}
L_{k,j}=\log\frac{\sum_{\bs_i\in\mathcal{X}^{0}_k}f_{\bY|\bX}(\by|\bs_i)}{\sum_{\bs_i\in\mathcal{X}^{1}_k}f_{\bY|\bX}(\by|\bs_i)},
\end{equation}
where $y$ is received symbols, $\bs_i$ denote the $i$th constellation point, $\mathcal{X}^{b}_k$ is the set of constellation symbols labeled by a bit $b\in\{0,1\}$ at bit position $k$.

In this paper, we assume the channel low  $f_{\bY|\bX}(\by|\bs_i)$ to be an $N$-dimensional Gaussian
distribution in \cite{ErikssonJLT2016,SillekensJLT2016},
\begin{equation}\label{eq:pdf}
\vspace{-0.5em}
f_{\bY|\bX}(\by|\bs_i)=\frac{1}{\sqrt{(2\pi)^N|\sum_{i}|}}e^{-\frac{1}{2}(\by_j-\bs_i)^{T}\bm{\sum}_i^{-1}(\by_j-\bs_i)},
\end{equation}
where $|\bm{\sum}_i|$ is the determinant of the covariance matrix $\bm{\sum}_i$ for  $i$th constellation point. In this paper, we consider two different assumptions of Gaussian distribution. In the first case, we assume independent and identically distributed Gaussian (4D-iidG model) random variables, and thus, $\bm{\sum}_i$ is a $4\times 4$ identity matrix multiplied with the average noise variance $\sigma^2$,  which is  identical for all $M$ constellation points. In the second case, we do not make an assumption on the covariance matrix, and thus, we assume correlated $N$-dimensional Gaussian noise (4D-CG model). 
Note that LLRs mismatched to the channel or computed with some approximation result in a rate loss \cite{AlvaradoJLT2018}.
Different LLR correction
strategies can be used to further improve the rate as well as the
performance of the FEC decoder, i.e., via LLR post-processing \cite{AlvaradoECOC2016}.

\begin{figure}[!tb]
	\centering
	\includegraphics[width=0.5\textwidth]{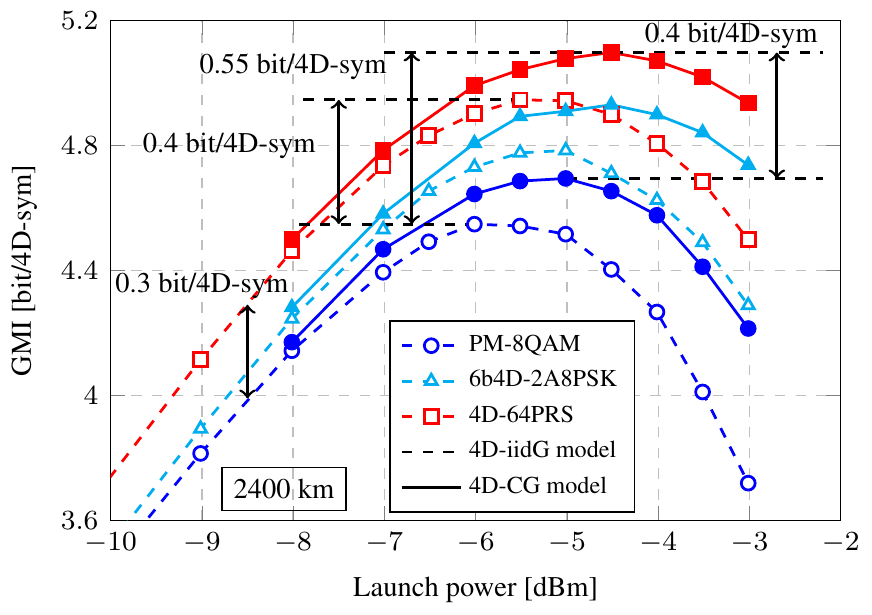}
	\vspace{-1.5em}
	\caption{Measured GMI versus launch power after 2400 km.}
	\label{fig:GMIvsP}
	\vspace{-0.5em}
\end{figure}

The transmission performance is shown in Fig.~\ref{fig:GMIvsP}  for three modulation formats (PM-8QAM, 6b4D-2A8PSK \cite{Kojima2017JLT} and 4D-64PRS) and two distribution models over  2400~km. In the linear regime, 4D-64PRS offers a GMI gain of 0.3 bit/4D-sym over PM-8QAM. This gain increases to 0.4 bit/4D-sym at the optimum launch power in the nonlinear regime. 
By using the 4D-CG model, the GMI improves by additional 0.15 bit/4D-sym compared to the 4D-iidG model. The largest obtained gain is therefore 0.55 bit/4D-sym gain ($12\%$ net data rate increase). Fig.~\ref{fig:GMIvsP} also shows that 4D-64PRS is more tolerant to nonlinearities, which is shown by the higher optimum launch power, and by the larger gain in the highly nonlinear regime.

\begin{figure}[!tb]
\centering
\scalebox{0.95}{    
     \begin{subfigure}[PM-8QAM]{
     	\includegraphics[width=0.5\textwidth]{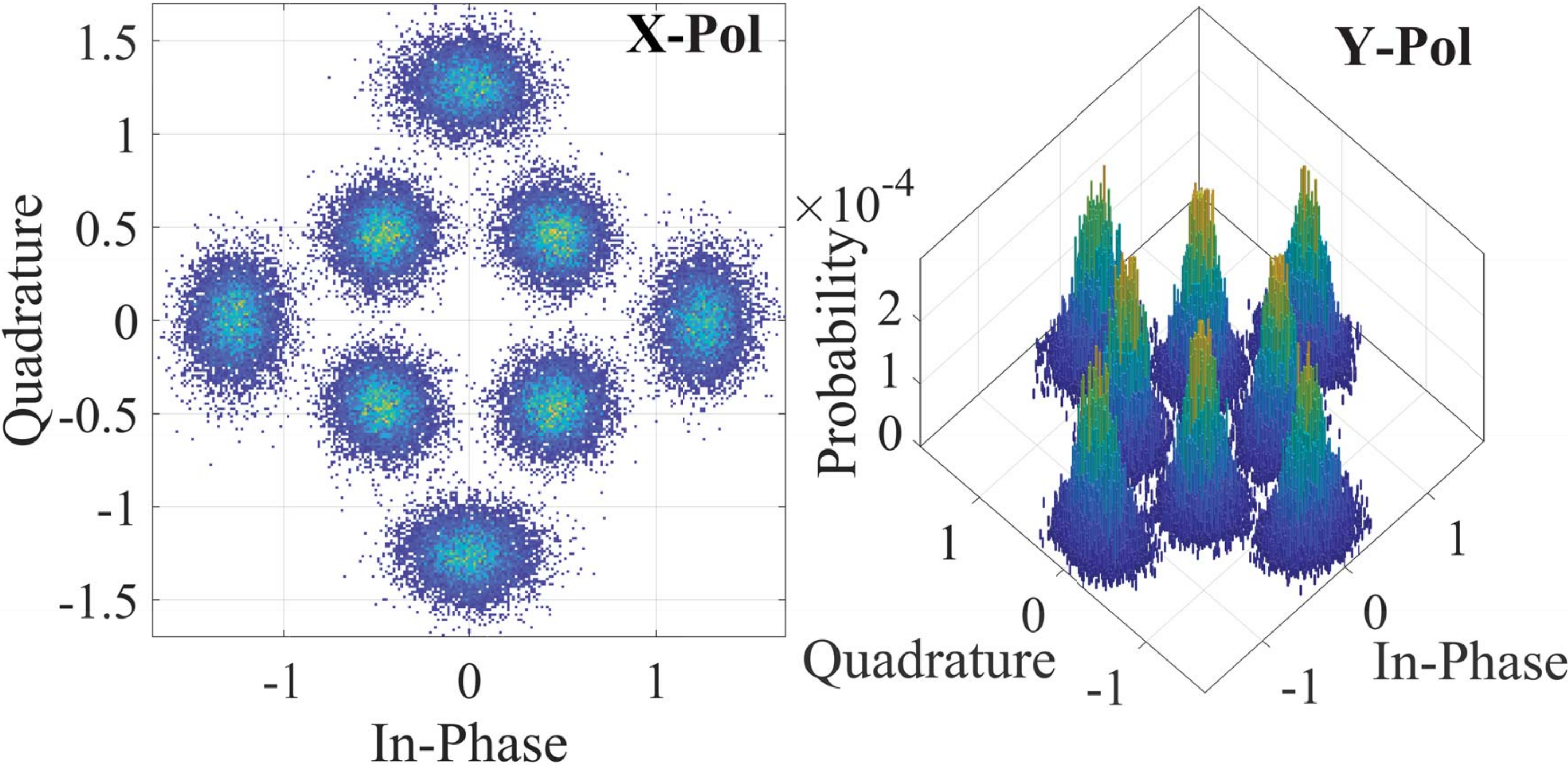}
     }
   \end{subfigure}
   }
   \\
\scalebox{0.95}{ 
    \begin{subfigure}[4D-64PRS]{
\includegraphics[width=0.5\textwidth]{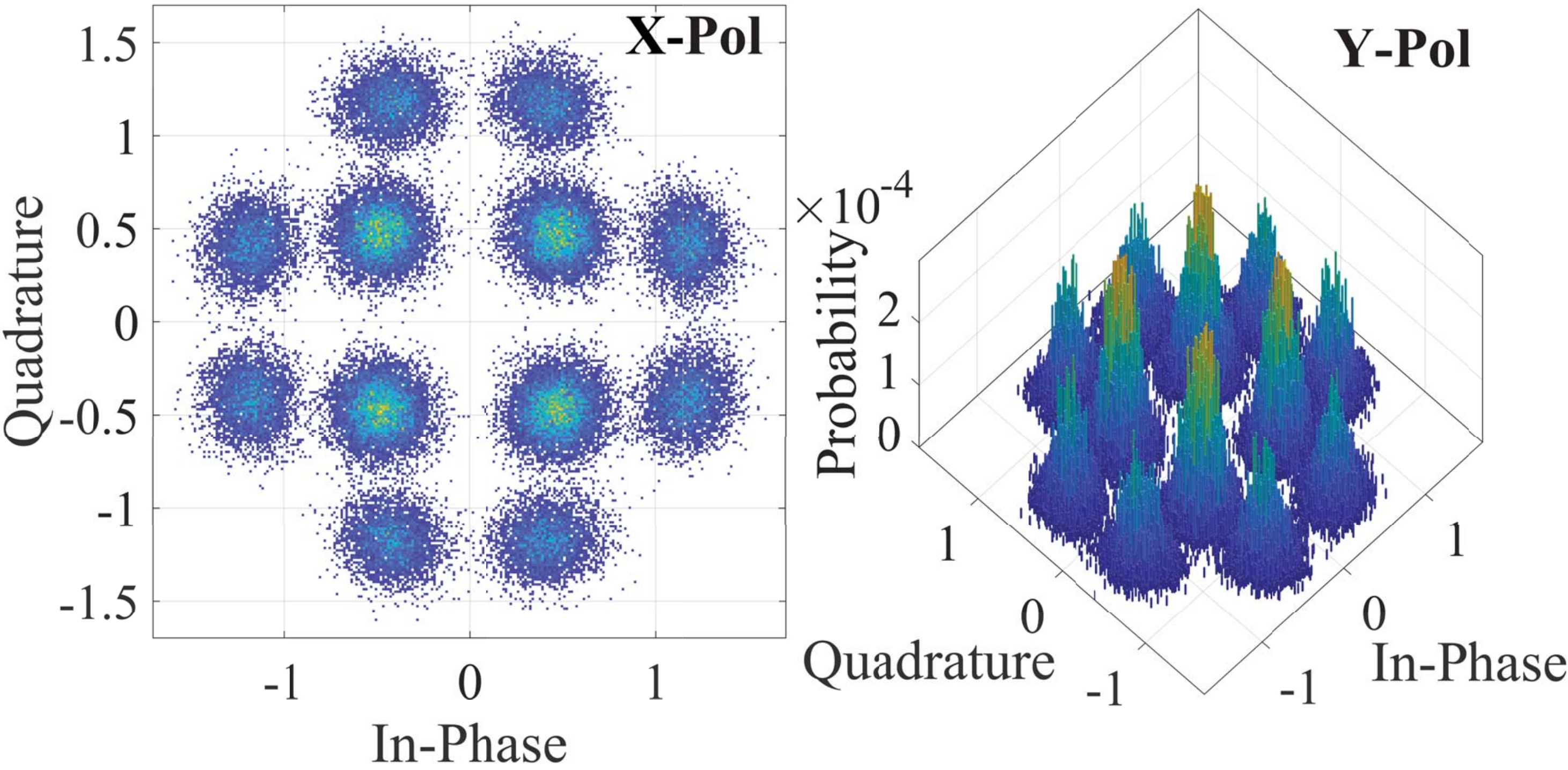}
     }
\end{subfigure}
}
\caption{Recovered constellations: (a) PM-8QAM and (b) 4D-64PRS for two  polarisations after 10 spans. The x-polarisation is displayed as a 2D projection, whilst the y-polarisation is shown as a 3D projection including the  probabilities.}
\vspace{-0.5em}
	\label{fig:ReceivedSymbols}
\end{figure}

Fig.~\ref{fig:ReceivedSymbols} shows the recovered constellations for PM-8QAM and 4D-64PRS at 800 km and optimum launch power (2D projection). 
The phase noise nature of NLIN is  visible in the case of PM-8QAM, but its relative significance reduces for 4D-64PRS. In addition, the noise distribution of 4D-64PRS is closer to a circularly-symmetric distributions. From Fig.~\ref{fig:ReceivedSymbols}, it is clear that the  NLIN partly depends on the modulation format and can be  mitigated by the constant-modulus property of 4D-64PRS. Compared to uniform PM-8QAM, Fig.~\ref{fig:ReceivedSymbols} also shows that 4D-64PRS induces a nonuniform distribution when projected onto 2D. 

\begin{figure}[!tb]
	\centering
	\scalebox{1.0}{\includegraphics[width=0.5\textwidth]{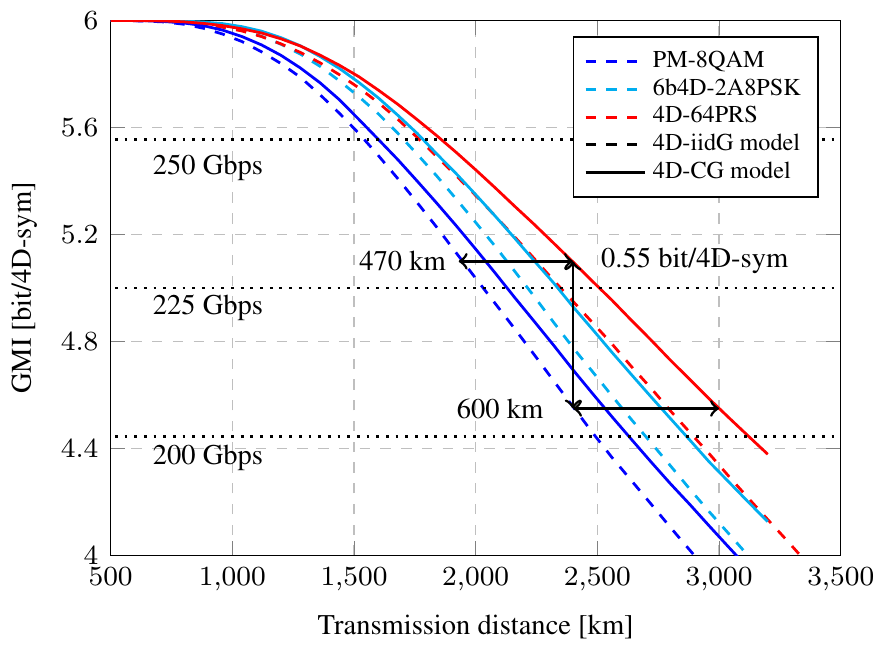}}
	\vspace{-1.5em}
	\caption{The measured GMI with respect to the transmission distances for 11 WDM channels at 250 Gbps, 225 Gbps and 200 Gbps achievable data rates.}
	\label{fig:GMIvsD}
\end{figure}

The GMI of PM-8QAM, 6b4D-2A8PSK and 4D-64PRS are compared in Fig.~\ref{fig:GMIvsD}. GMI of targeted three data rates shown in the dotted lines, which correspond to 250~Gb/s, 225~Gb/s and 200~Gb/s. For the 4D-iidG model under consideration, the 4D-64PRS is capable of outperforming PM-8QAM and 6b4D-2A8PSK at all three rates.
After applying 4D-CG model on the 4D-64PRS to achieve the same GMI of 4.55 bit/4D-sym, 4D-64PRS with 4D-CG model yields a reach increase of 600 km ($25\%$) and 500 km ($20\%$) with respect to PM-8QAM with 4D-iidG model and 4D-CG model, respectively. The transmission distances achieved by 4D-64PRS at the three rates are 1900~km,  2500~km and 3000~km, respectively.

\begin{figure}[!tb]
	\centering
	\scalebox{1.0}{\includegraphics[width=0.5\textwidth]{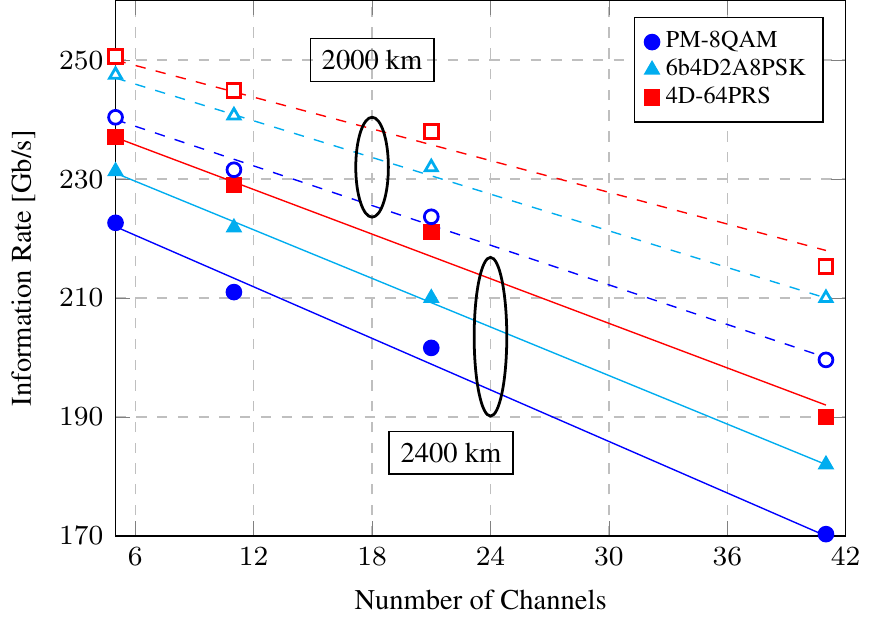}}
	\vspace{-1.5em}
	\caption{Dependence of the information rate at optimum launch power on the number of channels by using 4D-CG model. The solid/dashed lines are a linear fit of the simulation results.
}
	\label{fig:GMIvsNchannel}
	\vspace{-0.5em}
\end{figure}

To conclude, Fig.~\ref{fig:GMIvsNchannel} shows the dependence of the information rates (in Gb/s) on the number of channels at optimum launch power.  For 2400~km, 4D-64PRS yields an increase in information rate of 10 Gb/s ($4\%$) compared to PM-8QAM for the 5 channels case. This advantage increases with the number of channels,  which provides a gain of 20 Gb/s ($12\%$) for 41 channels. Similar gains are also observed at a distance of 2000~km.
This clearly shows the effectiveness of 4D-64PRS in suppressing nonlinearity.
\vspace{-0.5em}
\section{Conclusions}
\vspace{-0.5em}
We investigated the performance of the recently introduced 4D-64PRS modulation format in dispersion managed links. The nonlinearity tolerance of  4D-64PRS was shown to yield large reach and data rate increases with respect to regular PM-8QAM. These gains were shown to becomes even higher when the number of WDM channels increases. In this paper we also highlighted the importance of using the correct channel model assumption in the demapper, which resulted in additional gains.
We believe 4D-64PRS is a promising candidate for upgrading installed legacy low-dispersion fibres with high nonlinearity. The design of higher dimensional modulation formats and low-complexity demappers are left for future research.\\

\vspace{-0.8em}
\small
{
\noindent \textbf{Acknowledgements:}  This work was supported by Huawei France through the NLCAP project. The work of B. Chen is partially supported by the National Natural Science Foundation
of China (NSFC) under Grant 61701155. The work of A. Alvarado is supported by the Netherlands Organisation for Scientific Research (NWO) via the VIDI Grant ICONIC (project number 15685). The authors would like to thank Tobias Fehenberger and Eric Sillekens for useful discussion.}
\newpage

\balance
\section*{References}
\bibliographystyle{ECOC}
\bibliography{references}

\end{document}